\documentclass[a4paper,12pt]{article}
\newcommand{\pdiff}[2]{\ensuremath{\frac{\partial #1}{\partial #2}}} 
\newcommand{\phia}{\ensuremath{\phi_\alpha}}
\usepackage[english]{babel}
\usepackage[utf8x]{inputenc}
\usepackage{hyperref}
\usepackage{graphicx}
\usepackage{caption}
\usepackage[labelformat=simple]{subcaption}

\usepackage{mathtools}
\usepackage{xfrac}
\usepackage{siunitx}
\usepackage{placeins}
\usepackage{tikz}
\usepackage{pgfplots}
\usetikzlibrary{calc}
\usetikzlibrary{backgrounds}
\usepgflibrary{shapes}
\usetikzlibrary{through}
\usetikzlibrary{intersections}
\usetikzlibrary{matrix}
\usepackage{multirow}
\usepackage{hyperref}
\usepackage{cleveref}
\crefname{figure}{Fig.}{Figs.}
\Crefname{figure}{Figure}{Figures}

\crefname{table}{Table}{Tables}
\Crefname{table}{Table}{Tables}

\usepackage{xifthen}

\usepackage{esdiff}
\usepackage{ifthen}
\usepackage{authblk}

\usepackage{booktabs}
\usepackage{nicefrac}

\usepackage[
]{todonotes}
\presetkeys{todonotes}{inline}{}
\usepackage{footmisc}

\usepackage[singlespacing]{setspace}

\usepackage{soul}
\sethlcolor{black}
\makeatletter
\newif\if@blind
\@blindtrue
 \if@blind \sethlcolor{black}\else
    \let\hl\relax
 \fi
\renewcommand{\hl}[1]{#1}

\newcommand{\rz}{{\if mm {\rm I}\mkern -3mu{\rm R}\else \leavevmode
\hbox{I}\kern -.17em \hbox{R} \fi}}
\newcommand{\nz}{{\if mm {\rm I}\mkern -3mu{\rm N}\else \leavevmode
\hbox{I}\kern -.17em \hbox{N} \fi}}

\newcommand{\Vap}{\ensuremath{\mathrm{V}}}

\newcommand{\grad}[1]{\ensuremath{\nabla{#1}}}

\newcommand{\vv}[1]{\boldsymbol{#1}}

\newcommand{\ha}{\ensuremath{h_\alpha}}

\newcommand{\pace}{{\textsc{Pace3D}}}

\newdimen\CdotAxis
\newcommand*{\CdotAux}[3]{%
  {%
    \settoheight\CdotAxis{$#2\vcenter{}$}%
    \sbox0{%
      \raisebox\CdotAxis{%
        \scalebox{#1}{%
          \raisebox{-\CdotAxis}{%
            $\mathsurround=0pt #2#3$%
          }%
        }%
      }%
    }%
    \dp0=0pt %
    \sbox2{$#2\bullet$}%
    \ifdim\ht2<\ht0 %
      \ht0=\ht2 %
    \fi
    \sbox2{$\mathsurround=0pt #2#3$}%
    \hbox to \wd2{\hss\usebox{0}\hss}%
  }%
}

\def\addlegendimage{\csname pgfplots@addlegendimage\endcsname}

\DeclareSIUnit{\molpc}{mol\text{-}\%}

\title{An improved grand-potential phase-field model of solid-state sintering for many particles}
 \author[1*]{{Marco Seiz}}
 \author[1]{{Henrik Hierl}}
 \author[1,2]{{Britta Nestler}}
 \affil[1]{{Institute of Applied Materials, Karlsruhe Institute of Technology, Stra\ss{}e am Forum 7, 76131 Karlsruhe, Germany}}
 \affil[2]{{Institute of Digital Materials,  Karlsruhe University of Applied Sciences, Moltkestr. 30, 76133 Karlsruhe, Germany}}
 \affil[*]{{corresponding author: marco.seiz@kit.edu}}

\begin{document}

\maketitle


\begin{abstract}
Understanding the microstuctural evolution during the sintering process is of high relevance as it is a key part in many industrial manufacturing processes.
Simulations are one avenue to achieve this understanding, especially field-resolved methods such as the phase-field method.
Recent papers have shown several weaknesses in the most common phase-field model of sintering, which the present paper aims to ameliorate.
The observed weaknesses are shortly recounted, followed by presenting model variations aiming to remove these deficiencies.
The models are tested in the classical two-particle geometry, with the most promising model being run on large-scale three-dimensional packings to determine representative volume elements.
A densification that is strongly dependent on the packing size is observed, which suggests that the model requires further improvement.

\end{abstract}

\section{Introduction}

Sintering is an important processing step in the manufacturing of not only common goods such as coffee cups but also more specialized applications such as spark plugs in cars as well as the manufacturing of solar cells.
Furthermore, it is a natural process occurring on glaciers and thus plays a role in predicting avalanches as well as the evolution of the Earth's ice caps.
During sintering, the coupled processes of grain growth and densification take place.
The focus of this paper will rest on the latter.
Early theoretical models for densification concentrated on the geometric evolution of two particles evolving under diffusive conditions, which was then extrapolated to the entire sintering process.
This was found not to represent experimental data and thus separate models for the other stages --- in total initial, intermediate and final --- were developed.
While these captured the qualitative aspects of the sintering process quite well, for accurate quantitative predictions their model parameters often need to be fitted to experimental results.
Thus field-resolved methods such as the Monte-Carlo method or the phase-field (PF) method are adapted to model the sintering process, in which only simple material parameters enter.
These evolve the microstructure as a separate field, which allows for a 4D investigation of the process and eliminates the need for any geometric assumptions.
Recently it has been shown\cite{Seiz2022} that PF models of solid-state sintering need to include advective terms in order to reduce size effects on densification.
However, several problems of the most common method to calculate advective velocities\cite{Wang2006} were shown in the same publication.
Furthermore, the most common energy functional employed for the process was found to spontaneously generate voids on multi-grain junctions\cite{Yang2022b}.
Thus the goal of the present paper is to present a model reducing or outright eliminating these defects and apply it to find representative volume elements for sintering.
First, the model and its improvements will be detailed.
Following this, the model is investigated by running simulations and testing whether the problems found in prior simulations are eliminated.
Finally, large-scale three-dimensional simulations are conducted in order to find representative volume elements for sintering, in particular the densification of particle compacts.

%

\section{Modeling \& Methods}
In this section, the general model and its improvements will be described.
Furthermore, the employed analysis techniques will be detailed.

\subsection{General model formulation}
The model in the following is based on \cite{Hoetzer2019}, but extended with advection terms, which were found to be necessary to have almost constant densification as the green body size is varied\cite{Seiz2022}:

\begin{align}
\tau \epsilon \pdiff{\phi_\alpha}{t} + \nabla \cdot (\vec{v}_\alpha(x)\phi_\alpha) &= 
-\epsilon  \left(\pdiff{a(\phi,\nabla\phi)}{\phia} - \nabla \cdot \frac{\partial a(\phi,\nabla\phi)}{\partial\nabla\phia} \right) \\
&-\frac{1}{\epsilon}  \pdiff{w(\phi)}{\phia} \nonumber
- \sum^N_{\beta=1} \psi_{\beta}(\vv\mu,T) \pdiff{h_{\beta}(\phi)}{\phia} - \lambda \label{eq:phi}\\
\pdiff{\mu}{t} &=\left[ \sum_{\alpha=1}^N \ha(\phi) \left(\frac{\partial c_\alpha(\mu, T)}{\partial \mu} \right)\right]^{-1}  \\
&\Biggl( \nabla \cdot \Big(M(\phi,\mu,T)\grad{\mu} - \vec{v}(x)c \Big) 
- \sum_{\alpha=1}^N c_\alpha(\mu,T) \frac{\partial h_\alpha(\phi)}{\partial t} \Biggr). \label{eq:mu} 
\end{align}
These two partial differential equations describe the evolution of $N$ phase-fields as well as $K$ chemical potentials.
For the case of sintering, a surrounding vapor or vacuum phase $\alpha=0=\Vap$ is distinguished from solid grains of arbitrary orientation $\alpha \geq 1$.
Two components ($K=2$) are considered, namely copper and vacancies, with copper taken to be the independent concentration $c$.
Further details of the model terms and how the parameters affect the evolution equations are described in \cite{Hoetzer2019}, hence in the following only the modifications will be detailed.
First, as mentioned above, the chemical system now consists of copper and vacancies.
The free energy of solid copper ($s$) and the surrounding vapor (\Vap) are modelled with the standard parabolic approach 
\begin{align}
 G_\Vap(c,T) &= A_\Vap(T)(c-c_{\Vap}(T))^2 \\
 G_s(c,T) &= A_s(T)(c-c_{s}(T))^2
\end{align}
but with an assumption of stoichiometry, i.e. $A_{\Vap,s}(T) = A_{\Vap,s} = 50$ with the value being sufficiently high as to reduce the spontaneous shrinkage of grains within the phase-field context\cite{Yue2007}.
The grand chemical potential $\psi$ for each phase can then be obtained with a Legendre transform.
The equilibrium concentrations are arbitrarily assumed to be $c_s = 0.98$ and $c_\Vap=0.02$.
Combining this with setting the initial concentrations within the phases to be their equilibrium concentration, volume conservation between the two phases is largely achieved.
Furthermore, due to the equilibrium being independent of temperature and the flat equilibrium values being set initially, curvature becomes the main driving force, as expected from sintering.
The absolute level of the concentrations does not play a significant role.

The advection-free model can be shown to recover the Gibbs-Thomson condition via a thin-interface analysis\cite{Hoffrogge2021}, with the chemical driving force being decoupled from the surface energy\cite{Plapp2011}.
As these are decoupled, spontaneous void formation requires that the chemical driving force, represented by the grand potentials, imply void formation.
Considering the free energies above, this is only possible if the vacancy concentration is severely increased at multi-grain junctions.
However, the present model excludes the possibility of vacancy enrichment at grain boundaries (GBs) as there is no driving force for this.
Thus the model should be free of the spontaneous void formation observed by \cite{Yang2022b}.
Reproducing the test case of \cite{Yang2022b} indeed showed neither vacancy enrichment at GBs nor spontaneous void formation.

The interfacial diffusivities take into account the physical ($\delta_i$) and phase-field ($W$) interface widths by scaling these values as to match the diffusivity in the physical case:
\begin{align}
 \int_{-\delta_{i}/2}^{\delta_{i}/2} D_{i}^{real} dx &= \int_{-W/2}^{W/2} I(\phi_\alpha, \phi_\beta) D_{i}^{sim} dx \\
 \rightarrow D_{i}^{sim} &= D_{i}^{real} \frac{32\delta_{i}}{\epsilon \pi^2}
\end{align}
which is equivalent to \cite{Kundin2021a} except for the parametrization of the interface width.
The function $I(\phi_\alpha, \phi_\beta) = \phi_\alpha \phi_\beta$ interpolates the interfacial diffusion across the variable phase-field.
The input values as listed in the later table always describe $D_{i}^{real}$ and are transformed to $D_{i}^{sim}$ on simulation start.

The calculation of the grain velocities is based on the model outlined by Wang\cite{Wang2006}, but with a few modifications in order to account for the problems observed in \cite{Seiz2022}, which will be detailed in \cref{sec:improvements}.
Grain boundaries are assumed to act as vacancy sinks and sources.
When active, these induce a force $F_\alpha$ on their neighboring grains due to vacancy absorption or generation
\begin{align}
  F_{\alpha} &= \int_V dF_\alpha dV \\
  dF_{\alpha} &= \tilde{\kappa} \sum_{\beta \neq \alpha} (c-c_{gb})g(\alpha,\beta) (\nabla {\phi_\alpha} - \nabla {\phi_\beta})\label{eq:dF},\\
  g(\alpha,\beta) &= \begin{cases}
                  1, & \phi_\alpha \phi_\beta \geq \phi^{min}_{\alpha\beta}\\
                  0, & else\\
                \end{cases}.
\end{align}
The factors within the formulas are a stiffness $\tilde{\kappa}$, a grain boundary concentration $c_{gb}$, the switching function $g(\alpha,\beta)$, which identifies grain boundaries, and the phase-field gradients $\nabla \phi_\alpha$.
The gradient construction ensures that action and reaction are balanced, thus satisfying conservation of momentum.
The force is then assumed to result in an instantaneous velocity $v_\alpha$ via
\begin{align}
  V_\alpha &= \int_V \phi_\alpha dV \\
  v_\alpha &= m_t\frac{F_\alpha}{V_\alpha}
\end{align}
with the volume $V$ accounting for particle size and the factor $m_t$ representing a translational mobility.
A torque would be generated by an asymmetric distribution of $dF_{\alpha}$ across grain boundaries, but since these were shown not to have an influence of densification\cite{Shi2021} these terms are dropped.
Since the factors $\tilde{\kappa}$ and $m_t$ only appear together, they are melded to a single factor $\kappa = \tilde{\kappa}m_t$, which will henceforth be called the effective stiffness.
Finally, the phase velocities $v_\alpha$ are interpolated using local phase-fields:
\begin{align}
 v_\alpha(x) &= v_\alpha \phi_\alpha(x) \label{eq:vinterp}\\
 v(x) &= \sum v_\alpha(x),
\end{align}
with $v_\alpha(x)$  being used to locally advect each grain phase-field $\alpha$.
The velocity $v(x)$ is used to advect the surrounding vapor as well as the concentration.

\subsection{Model improvements}
\label{sec:improvements}
The first improvement concerns the factor $c_{gb}$ in \cref{eq:dF}.
In \cite{Seiz2022} it was observed that if $c_{gb}$ deviates from the ``true'' equilibrium concentration $c^{eq}_{gb}$ on a grain boundary, then two problems can occur:
First is the problem of ``unsintering'', i.e. the system densifies up to a certain point after which the grain velocities force it apart, which also increases the system's free energy.
This happens if $c_{gb}$ is chosen to be smaller than $c^{eq}_{gb}$, which depends on the simulation state via the particle's curvatures.
Second, any choice of $c_{gb}$ which deviates from $c^{eq}_{gb}$ will lead to a different dihedral angle being observed.
This is due to the advective fluxes not vanishing when the dihedral angle as predicted by Young's law is achieved.
Thus the choice of $c_{gb}$ is indeed critical.
In essence, $c_{gb}$ needs to be chosen to be consistent with the equilibrium state defined by the energy functional --- otherwise, a minimization of free energy is no longer guaranteed and the supposed equilibrium properties of the model without advection need not hold.
Both problems vanish if $c_{gb} = c_{gb}(S) = c^{eq}_{gb}$ is achieved for the entire simulation run given that the simulation state $S$ allows the calculation of $c^{eq}_{gb}$.
Two methods for approximating this will be explored in the present paper.

Both methods are based on the relationship
\begin{align}
  c^{eq}_{gb} &= c_\alpha(\mu_{eq}+\Delta \mu, T)\label{eq:ceq-calc}
\end{align}
which assumes that $c^{eq}_{gb}$ is equivalent to the bulk concentration of an $\alpha$ grain plus a deviation in the chemical potential $\Delta \mu$.
The function $c_\alpha(\mu,T)$ is thermodynamically defined via $c_\alpha(\mu,T) = \pdiff{\psi_\alpha(\mu, T)}{\mu}$.
The methods then only differ in how $\Delta \mu$ is estimated.
The first method is based on the Gibbs-Thomson equation
\begin{align}
  \Delta \mu = \gamma\kappa\label{eq:gibbsthomson},
\end{align}
which describes the change of chemical potential $\Delta \mu$ from a planar surface to a curved surface, employing the surface energy $\gamma$ and the curvature $\kappa$.
While $\gamma$ is known as the input parameter, the curvature $\kappa$ needs to be estimated from the current simulation state.
This can be done with a shape assumption, i.e. a circle (2D) or a sphere (3D), based on which an estimate for the radius of curvature can be easily obtained:
\begin{align}
 \Delta \mu_{2D} &= \gamma (\frac{V_\alpha}{\pi})^{-1/2} \label{eq:muV}\\
 \Delta \mu_{3D} &= \gamma (\frac{3V_\alpha}{4\pi})^{-1/3}.
\end{align}

Alternatively the discrete curvature $\kappa = \nabla \cdot \vec{n}$ could be calculated and then input into \cref{eq:gibbsthomson}.
However, this necessarily includes some cut-off, as the curvature is not well-defined close to the bulk regions\cite{Vakili2017}.
This problem can be avoided, while still accounting for the curvature, by observing the chemical potential on the surface.
Assume that the average chemical potential
\begin{align}
  \hat{\mu}_\alpha = \frac{\int \mu \phi_\alpha \phi_\Vap dV}{\int \phi_\alpha \phi_\Vap dV} \label{eq:muhat}
\end{align}
on a particle's surface gives an approximation to the chemical potential in equilibrium including the effect of \cref{eq:gibbsthomson}.
Then the deviation from a planar surface $\Delta \mu$ is equivalent to  $\hat{\mu}_\alpha$.
If the equilibrium chemical potential $\mu_{eq}$ of a plane surface is nonzero, subtract it from $\hat{\mu}_\alpha$.
This also naturally includes the effects of variable surface energies between interfaces.
Grains which do not have an interface with the vapor are assumed to have $\Delta \mu = 0$.
This ensures that a grain boundary at equilibrium will not be moved out of equilibrium with advection.
In both cases, this yields estimates for $\Delta \mu_\alpha$ for each grain phase $\alpha$.
The value employed within an $\alpha\beta$ grain boundary will then simply be the average of both.

The second improvement concerns the interpolation of velocities, which might not seem significant at first, but it can in fact lead to rarefaction and shocks on the interface.
These can change the equilibrium profile significantly and thus the effective surface energy.
Consider the classical interpolation described above \cref{eq:vinterp}, and only the advection part of the phase-field equation,
\begin{align}
\vec{v}_\alpha(x) &= \vec{v}_\alpha \phi_\alpha(x)\\
\pdiff{\phi_\alpha}{t} &= - \nabla \cdot (\vec{v}_\alpha(x)\phi_\alpha(x))\\
&= - \nabla \cdot (\vec{v}_\alpha \phi_\alpha(x)^2)
\end{align}
i.e. the flux due to ``advection'' is now quadratic in the advected variable.
However, this is precisely the inviscid Burgers equation up to a multiplicative constant $\vec{v}_\alpha$ and thus rarefaction and shocks will naturally occur.
Since there is no physical reason for either during sintering, this type of equation needs to be avoided.
A simple way of avoiding this effect is to always advect the grains with their actual rigid-body velocity $\vec{v}_\alpha$.
However, the surrounding vapor as well as the concentration field still need to be advected.
For these, a formulation which largely avoids this problem is
\begin{align}
\vec{v}_c(x) &= \frac{\sum_{\alpha != v}\vec{v}_\alpha \phi_\alpha(x)} {\sum_{\alpha != v} \phi_\alpha(x)}
\end{align}
which yields a jump function across a grain-vapor interface for an obstacle-type potential.
This avoids rarefaction to a large extent, as a constant velocity is seen on grain-vapor interfaces right up to the bulk region of vapor.
Shocks can possibly form on the transition from the interface to bulk vapor, but since this is concentrated on a small part of the interface, the effect is negligible.
Across a grain-grain interface the concentration equation still has more of a Burgers like character.
However, the actual flux is small as the concentrations are close to each other, with no rarefaction or shocks being observed across grain-grain interfaces in the simulations.
It is noted that \cite{biswasimplementation} employed a similar strategy in order to ``boost the numerical convergence of the model'', though without explicitly identifying the Burgers-like character of the original equations.

Finally, the force density $dF_\alpha$ of \cref{eq:dF} is weighted by the grain boundary ``phase'' $\sum_{\beta} 4\phi_\alpha\phi_\beta$ with $\beta \notin \{\alpha, \Vap\}$, and later divided by the integral of the same quantity.
This makes the jumps in velocity as observed in \cite{Seiz2022} less egregious, but has little qualitative influence on the results.

\subsection{Computational aspects}
The solver for the model is implemented with finite differences within the {\pace{}} framework {\cite{Hoetzer2018}}, based on the massively parallel and high-performance implementation of {\cite{SC19}}.
A WENO-5 scheme\cite{Shu2009} is employed for the calculation of the advection updates in order to reduce numerical diffusion.

The employed nondimensionalization scales are listed in \cref{tab:nondim} and the material parameters in \cref{tab:params}.
The grid spacing $\Delta x$ will be repeatedly varied and thus will be mentioned for each set of simulations.
The time step is calculated by estimating the stable time step in the explicit time integration scheme as well as the Courant-Friedrichs-Lewy (CFL) condition, with a safety factor of 0.3:
\begin{align}
 \Delta t = 0.3 \min(\Delta t_{\phi,c}, \Delta t_{CFL})\nonumber\\
 \Delta t_{\phi,c} = \frac{\Delta x^2}{2 \max(D_{\phi}, D_{c})}\nonumber\\
 \Delta t_{CFL} = \bigl[\sum_{i} \frac{\max(|v_i|)}{\Delta x_i}\bigr]^{-1}\nonumber
\end{align}
with the effective phase-field diffusivity $D_\phi = 2\frac{\max(\gamma)}{\min(\tau)}$ with the respective maximum and minimum values of $\gamma$ and $\tau$, the highest diffusivity employed for the concentration equation $D_c$, $i$ going over the spatial dimensions and $\max(|v_i|)$ being the largest velocity per dimension.
Typically though the phase-field step is the limiting factor for stable time integration.

The values of the interfacial energies are based on estimates for pure copper at \SI{700}{K}, resulting in a dihedral angle of $\SI{151}{\degree}$.
The grain boundary diffusion value is based on \cite{Suzuki2005}, with the surface diffusion value being based on \cite{Butrymowicz1973}.
The bulk diffusion within the grains and vapor will be varied, and thus be mentioned for each simulation set.
When employing \cite{Butrymowicz1973}, bulk Cu diffusion would be at $\SI{1e-20}{m^2/s}$ for \SI{700}{K}, effectively freezing the diffusion field within the bulk relative to the interfacial diffusivities.
Instead of using this tiny value, larger values will be used in order for allow a reasonable amount of relaxation within the grains; this should not significantly influence the qualitative results if faster diffusion mechanisms (grain boundary, surface) are active at the same time.
The kinetic coefficient of the surfaces $\tau_{v\alpha}$ is chosen such that the phase-field always relaxes more quickly than the chemical potential, which ensures that the process is controlled by diffusion.
Grain growth, if thermodynamically possibly, is largely suppressed by taking the kinetic coefficient between grains to be  $100\tau_{v\alpha}$.
The effective stiffness $\kappa$ is chosen based on the observations in \cite{Shi2021}, such that the simulation results become independent of its choice:
The advective velocity tends to increase as $\kappa$ is increased until a plateau is reached.
This plateau is determined in a pre-study and found to start at 800, with $\kappa=3200$ employed in the simulations to ensure that the results are independent of $\kappa$.
The resulting data and evaluation of this pre-study is available within the Supplementary Material.

This pre-study as well as the small scale validation in \cref{sec:vali} are calculated on a local machine using GNU parallel\cite{parallel} for effective job management.
The later large scale simulations are calculated on the Hawk supercomputer at the HLRS.

\begin{table}[h]
\centering
\caption{nondimensionalization parameters}
\label{tab:nondim}
 \begin{tabular}{ll}
scale   & value \\
length $l_0$ &  $\SI{1e-8}{m}$  \\
diffusivity $D_0$ & $\SI{1e-12}{m^2/s}$\\
time $t_0$ & $\SI{1e-4}{s}$ \\
velocity $v_0$ & $\SI{1e-4}{m/s}$\\
temperature $T_0$ & $\SI{700}{K}$ \\
surface energy $E_{s,0}$ & \SI{1}{J/m^2} \\
energy density $E_{r,0}$ & \SI{1e8}{J/m^3} \\
molar volume $V_{m,0}$ & \SI{7.1e-6}{m^3/mol} \\
\end{tabular}
\end{table}

\begin{table}[h]
\centering
\caption{Employed physical and numerical parameters for the simulations.}
\label{tab:params}
 \begin{tabular}{lll}

parameter   & nondim. value & physical value \\
\multicolumn{3}{c}{\textit{Numerical parameters}}\\
interface parameter $\epsilon$   &    $4\Delta x$  &   variable  \\
interface width $W$   &    $10\Delta x$  &   variable  \\
grain boundary cutoff $\phi^{min}_{\alpha\beta}$ & 0.14 & - \\
\multicolumn{3}{c}{\textit{Physical parameters}}\\
surface energy $\gamma_{v\alpha}$     & 2 & $\SI{2}{J/m^2}$ \\
grain boundary energy $\gamma_{\alpha\beta}$     & 1 & $\SI{1}{J/m^2}$ \\
grain boundary diffusion  $D_{gb}$    & 55               & $\SI{5.5e-11}{m^2/s}$  \\
surface diffusion  $D_s$ & 169               & $\SI{1.69e-10}{m^2/s}$   \\
physical interface width $\delta_i$ & 0.02 & $\SI{2e-10}{m}$ \\
surface kinetic coefficient $\tau_{v\alpha}$ & 0.08 & $\SI{8e10}{Js/m^4}$ \\
grain boundary kinetic coefficient $\tau_{\alpha\beta}$       & 100 $\tau_{v\alpha}$ & $\SI{8e12}{Js/m^4}$ \\
stiffness $\kappa$ & 3200 & - \\
\end{tabular}
\end{table}

\subsection{Data evaluation}
The primary variables of interest in the present paper are the following:
\begin{itemize}
 \item neck size $X$ between two particles
 \item dihedral angle $\theta$ between two particles
 \item total free energy $\mathcal{F}$ of the system
 \item linear shrinkage $e$ between two particles
 \item density $\rho$ of a packing
\end{itemize}

The neck size $X$ between two particles is assumed to be half the length of the grain boundary joining them.
The volume of the grain boundary is obtainable from the phase-field, but must be corrected by an interface profile dependent factor in order to obtain the grain boundary area:
\begin{align}
 V_{gb} &= \int_V 4\phi_\alpha\phi_\beta dV \\
 A_{gb} &= \frac{V_{gb}}{\int 4\phi_\alpha \phi_\beta dx}
\end{align}
Essentially, the volume needs to be divided by the amount of grain boundary ``phase'' $4\phi_\alpha\phi_\beta$ which occurs along the profile.
This is in general dependent on the local simulation state, as high driving forces can distort the interface.
However, in the present paper the profile should always be near equilibrium as only capillary forces and advection take place.
Thus the equilibrium solution for the profile can be employed, in which case $\int 4\phi_\alpha \phi_\beta dx = \frac{\pi^{2} \epsilon}{8}$.
In two dimensions, the grain boundary ``area'' calculated thus is actually the grain boundary length and no further geometric assumptions need to be employed.

The neck size is used in geometrical models of early stage sintering as a variable of interest\cite{Rahaman2003}.
These models generally predict a power law behavior $X/R = At^{1/n}$ with the initial particle radius $R$, the time $t$ and some constants $A, n$.
While $A$ depends on the material and geometrical properties, the constant $n$ should only depend on the dominant diffusion mechanism, with $n \in \{4, 5\}$ in the case of bulk diffusion\cite{Rahaman2003}.
Furthermore, the time evolution of properties such as the neck size should follow Herring's scaling law\cite{Herring1950}.
In effect, it says that if time is rescaled as $t/R^Z$, then relative properties such as the relative neck size should map back onto a master curve independent of particle size $R$.
The constant $Z$ depends on the dominant diffusion mechanism, with $Z=3$ for bulk diffusion.

The dihedral angle $\theta$ is calculated by fitting circles to parts of the 0.5 isoline of both particles.
The angle formed by the circles at their intersection, close to the triple point, is the dihedral angle.
After calculating the intersection point, the circles' individual angles are calculated via their slopes and subtracted from each other, yielding the total angle.
The relevant part of the isoline is that which contains the surrounding vapor, but does not contain the other particle i.e. excluding the flat grain boundary.

The linear strain between two particles is calculated by tracking the barycenters of the particles.
The linear strain follows as $e = \frac{L(t)-L(0)}{L(0)}$, with $L(t)$ describing the distance between the barycenters of the first and last particle as a function of time $t$.
This strain is also predicted to follow a power law $e = Bt^{2/n} = Bt^{1/m}$\cite{Rahaman2003}, with the exponent following from the neck growth law and $B$ being a different agglomeration of materials and geometrical parameters.
The strain will in general be negative (sample shortens) and as to allow easy fitting of the power law, the absolute value of this strain will be employed.
Based on the equilibrium dihedral angle $\theta$  and some geometric assumptions, \cite{B.J.Kellet1989} derived an expression for the equilibrium strain of an infinite chain of cylinders $|e_{eq|} = |1-R(\theta)cos(\theta/2)|$ with the equilibrium radius $R$.
For the parameters employed in this study $\theta = \SI{151}{\degree}$ and thus $|e_{eq}|=0.555$.

The density of a packing is determined by building a convex hull $C$ around the packing, employing the particles' barycenters as the point set to bound.
The density is then given by $\rho = \frac{\int_C \phi_{\alpha\geq1} dV}{\int_C \phi_\alpha dV}$, i.e. the ratio of solid phase-fraction within $C$ relative to the total volume of $C$.
A full sampling of the inner hull as well as a Monte Carlo (MC) approach similar to \cite{Greenquist2020} are tested.
The present MC approach uses the Gaussian Stopping Rule of \cite{Bicher2022}, with a confidence level of $p=95\%$ and a confidence interval width of $\delta_{abs}=0.005$ on the density.
The full sampling and the MC approach are compared for the smallest packings and found to have no large difference.
Since the MC sampling approach is faster, it will be employed henceforth.
It is noted that the convex hull includes significant amounts of space around the green body if the green body itself isn't convex yet; this is effectively the problem of the caliper measurement mentioned in \cite{Greenquist2020}.
This can lead to seemingly unphysical drops in density, since if any of the outermost particles move outwards, the convex hull gains a comparatively large amount of ``open'' porosity which is actually outside the green body proper.
Future work will consider employing concave hulls in order to sidestep this problem, but the convex hull approach suffices for the present.

\section{Results \& Discussion}
In this section the model without the improvements as well as the models with improvements will be compared.
The first simulation setup for this purpose is the venerable two-particle model, as it suffices to clarify whether the problems observed in \cite{Seiz2022} are fixed by the improvements or not.
The second setup concerns the scaling of the advective velocity with the green body size.
In \cite{Seiz2022} a small but persistent slowing of the densification speed with the green body size was observed, even with advective terms included.
This is explored by employing the most promising model from the two-particle setup in a three-dimensional packing and comparing it to a purely diffusive model.

\subsection{Equilibria and dynamics for two particles}
\label{sec:vali}
Two particles of equal radius $R$ are set symmetrically in a simulation box with no-flux conditions on all boundaries.
The box size is taken to be $4R + 9\epsilon$ in the direction where the particles touch, ensuring that the phase-field does not initially touch the boundary.
Directions perpendicular to this direction are of size $4R$, which is sufficient to ensure that the equilibrium states' phase-fields will not touch the boundary.
All phases are set to their equilibrium concentrations initially.

The following models will be considered for the present investigation:
A diffusion-only (DO) model, whose advective velocity is always zero.
Three models including advective terms (ADV), with the following variations:
\begin{itemize}
 \item a constant $c^{eq}_{gb}=0.99$ slightly above the equilibrium bulk concentration (C)
 \item estimating $c^{eq}_{gb}$ with the particle size (V) \cref{eq:muV}
 \item estimating $c^{eq}_{gb}$ with the average chemical potential on the surface ($\mu$) \cref{eq:muhat}
\end{itemize}
The DO model serves as a reliable baseline for the equilibrium shape, which the ADV models should match if they are consistent with the energy functional. 
The ADV (ADV-C, ADV-V, ADV-$\mu$)  models are expected to have faster neck growth and densification, with differences in their individual dynamics and possibly equilibrium states.
For simplicity of presentation, only the case of bulk diffusion will be considered, i.e. the coefficients for grain boundary and surface diffusion are set to zero.
The Cu diffusion in the grain is arbitrarily set to $D_b = \SI{1e-12}{m^2/s}$, with a value of $D_v = D_b/1000$ being used for the diffusion in the vapor.
The equilibrium properties will be independent of these choices for the DO model, while for the ADV models it will depend on whether they are consistent with the energy functional.
If these are not, then the choice of diffusion constants will influence the equilibrium.
The dynamic evolution will of course differ if the diffusion coefficient is changed, but the scaling with time will be the same.
Thus the qualitative aspects should readily transfer to cases with grain boundary or surface diffusion active as well as arbitrary non-zero choices of diffusion constants.

The first investigation is conducted at a constant particle size of $R=\SI{25}{nm}$, resolved with $r=25$ cells at $\Delta x = 0.1$.
An approximation for the chemical potential in equilibrium is given by $\mu_{eq} = \Delta \mu = \gamma_s\kappa$ and assuming $\kappa = 1/R_0$ with the initial radius $R_0 = 25\Delta x = 2.5$, which yields $\Delta \mu = 0.8$.
This can be translated into a bulk concentration by $c_\alpha(\mu) = c_s + \pdiff{c_\alpha}{\mu}\Delta \mu = 0.98 + \frac{0.8}{100} = 0.988$.

\begin{figure}
\centering
  \includegraphics[width=0.5\textwidth]{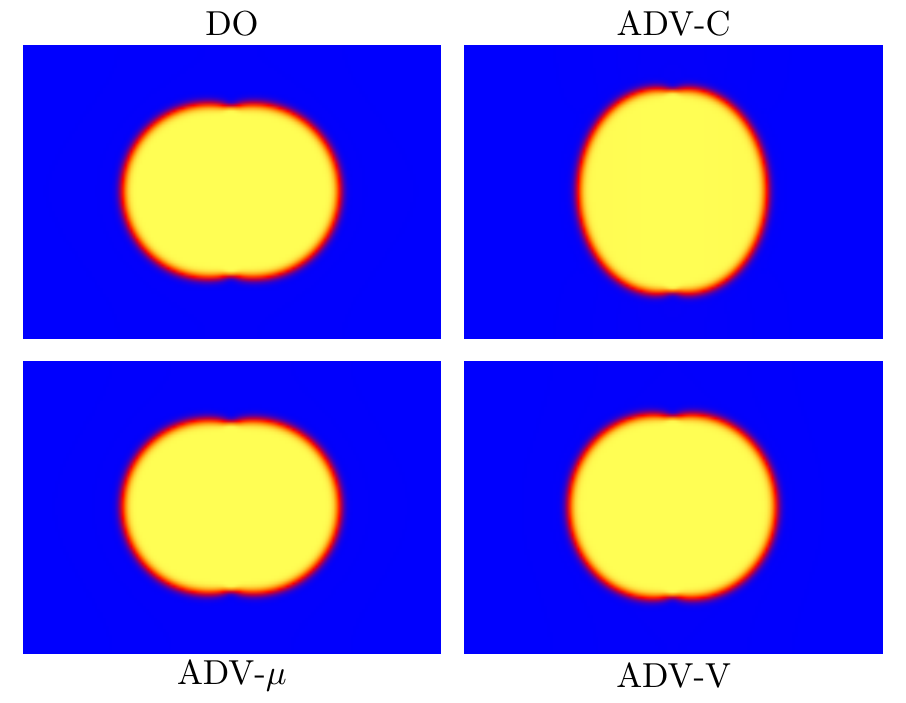}
  \caption{Obtained equilibrium shapes represented by the Cu concentration field, with yellow indicating the solid grains, blue the surrounding vapor and reddish-orange their interface.
  Note that model ADV-C results in a much more oblong shape, with remaining models showing similar shapes.}
  \label{fig:eqshapes}
\end{figure}

\begin{figure}[]
 \begin{subfigure}[]{0.45\textwidth}
     \includegraphics[width=\textwidth]{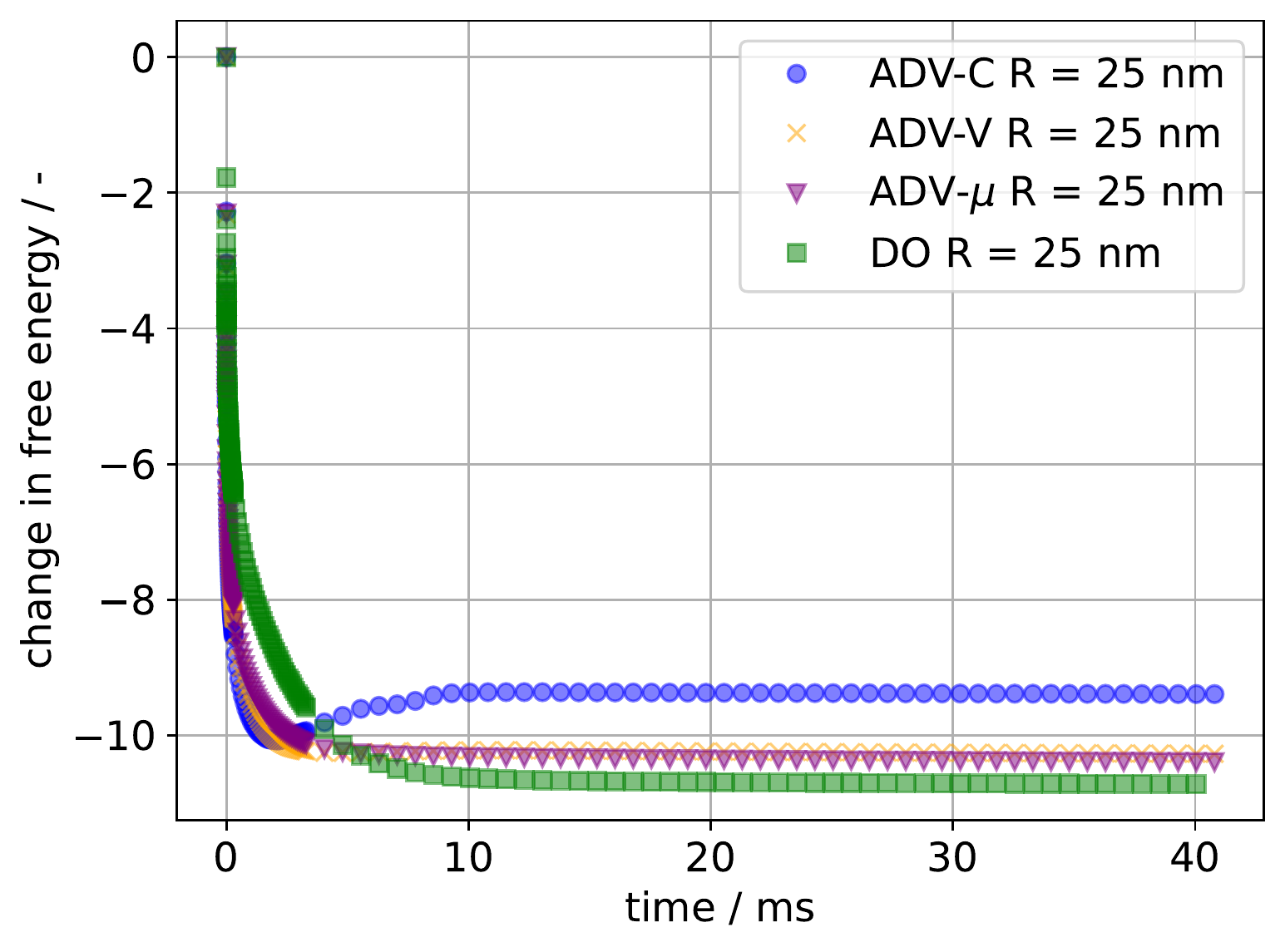}
    \caption{change of free energy
    }
    \label{fig:energy-time-r25}
  \end{subfigure}
  \begin{subfigure}[]{0.45\textwidth}
    \includegraphics[width=\textwidth]{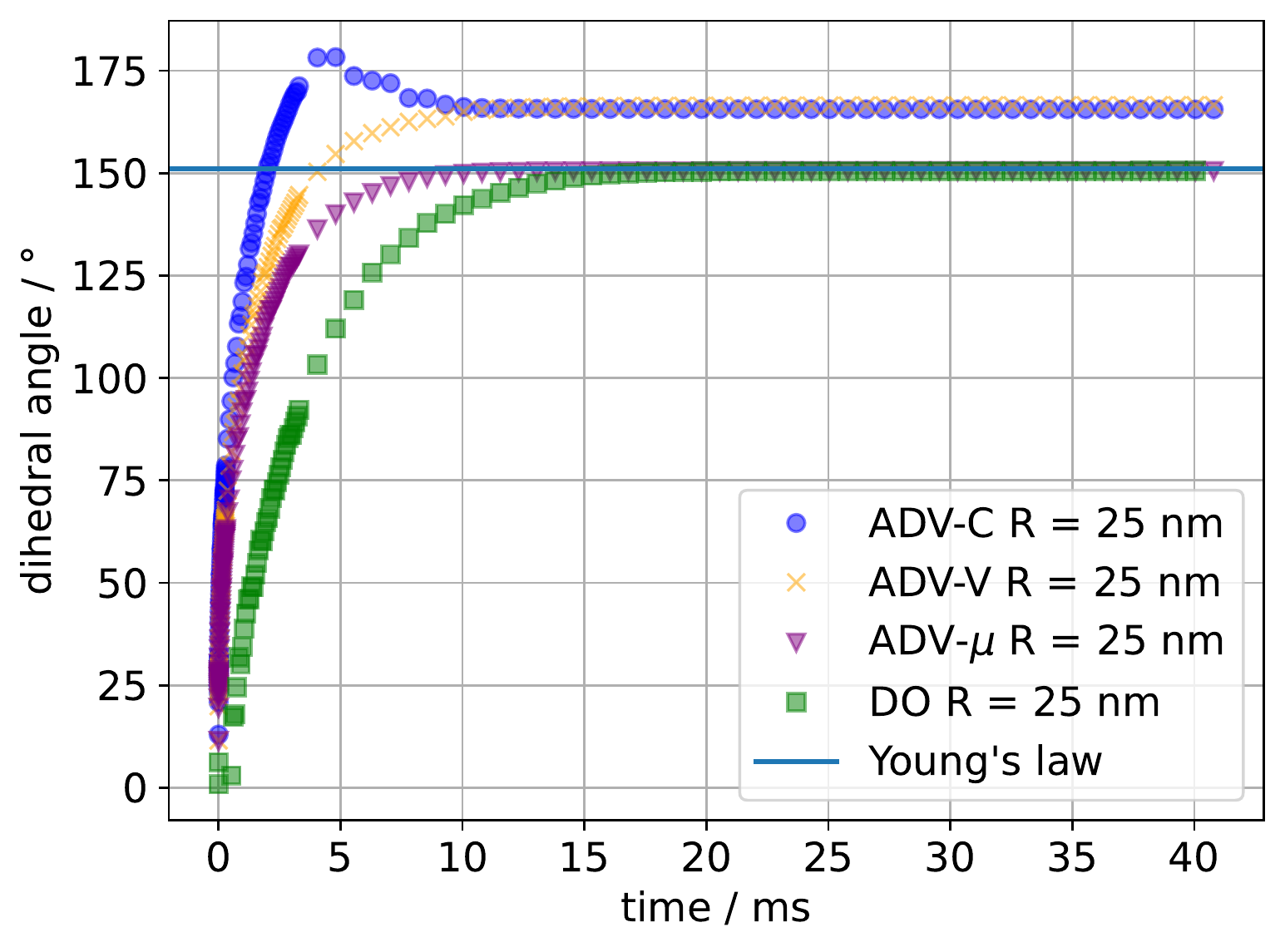}
    \caption{evolution of the dihedral angle
    }
    \label{fig:angle-time-r25}
  \end{subfigure}
  \caption{The models ADV-C and ADV-V show a non-monotonic evolution of the free energy, whereas models ADV-$\mu$ and DO show a monotonic drop in free energy.
  The theoretical dihedral angle is closely approximated by ADV-$\mu$ and DO as well, while models ADV-V and ADV-C significantly increase the angle.
  }
  \label{fig:en-angle-time}
\end{figure}

The simulations are continued until a state close to equilibrium is reached, with the obtained equilibrium shapes shown in \cref{fig:eqshapes}.
While models except for ADV-C show more or less similar equilibrium shapes, the shape of ADV-C is much more oblong due to its severe violation of minimization of free energy.
The free energy as well as dihedral angle will thus serve as tests on the consistency with the free energy functional.
The change in free energy is shown in \cref{fig:energy-time-r25}, relative to $t=0.0075$ in order to exclude the initial large jump from a sharp to a diffuse interface.
It is observed that the models except for ADV-C and ADV-V show a monotonic reduction in free energy.
For model ADV-V the non-monotonicity is short-lived and handily overshadowed by the other symbols, but simple forward differences showed that it also contains a non-monotonic reduction in free energy.
The observed equilibrium concentration (model DO) within the particles is about $0.9868$, which compares reasonably with the above simple approximation.
The remaining difference is easily explained, as multiple interfaces with different interface energies exist, which the estimate for $\Delta \mu$ doesn't take into account.

Although $c^{eq}_{gb}=0.99$ lies above this equilibrium concentration as suggested by \cite{Seiz2022}, an increase in free energy is observed.
As shown in \cite{Seiz2022}, the force density within the grain boundary region defined by $\phi_\alpha\phi_\beta > 0.14$ has repulsive (grain boundary) and attractive (triple point) regions.
During transient growth of the neck, the advective flux tends to decrease itself by lengthening the repelling grain boundary until it matches the diffusive flux.
Given that the diffusive flux acts densifying for dihedral angles below the equilibrium value, this implies that the advective flux has to increase the grain boundary length and thus dihedral angle beyond their equilibrium values in order to match the diffusive flux.
Models ADV-$\mu$ and ADV-V can potentially avoid this problem by decreasing the advective flux not by a grain boundary lengthening, but by decreasing the force density within the grain boundary.
The difference in free energies in equilibrium between models DO and ADV-$\mu$ are due to the spatially variable chemical potential field for ADV-$\mu$.
There is a finite, but small velocity remaining even for ADV-$\mu$ which balances out the diffusive flux within the grain boundary.

The dihedral angle $\theta$ is shown in \cref{fig:angle-time-r25}.
The DO and ADV-$\mu$ models achieve the same equilibrium dihedral angle $\theta = \SI{150.4}{\degree}$, missing the theoretical value by $\SI{0.6}{\degree}$.
However, both ADV-V and ADV-C increase the dihedral angle to about $\SI{166}{\degree}$.
As previously observed in \cite{Seiz2022}, the equilibrium dihedral angle is modified by a constant $c^{eq}_{gb}$ and thus this was to be expected.
At first glance, model ADV-V increasing the dihedral angle would seem odd, given that the simulation state is employed for estimating $c^{eq}_{gb}$.
However, the model for predicting $\Delta \mu$ assumed constant $\gamma$ for the interfaces, whereas in the simulation the surface and grain boundary energy are different.
This leads to a different equilibrium, which in the present case by happenstance is close to the ADV-C equilibrium.
It is likely that model ADV-V would perform much better for equal surface and grain boundary energy, but it seldom happens that these are equal.
In total, the only advective model that is observed to be consistent with the free energy functional is ADV-$\mu$.

All models will also be tested for adherence to Herring's scaling law.
For this, the radius $R$ will be varied by increasing the number of cells employed to resolve the particle $r$ as well as by changing the grid spacing $\Delta x$.
This is done as to verify that size effects have been fully included.
If the physical size $R = r\Delta xl_0$, with the nondimensionalization length $l_0$, is the same between two simulations with differing $\Delta x$, then similar curves should be obtained, with the difference entirely attributable to the  discretization error.
The number of cells employed to resolve the particle $r$ is in the set $\{25, 50, 100\}$, with two grid spacings $\Delta x \in \{ 0.1, 0.2 \}$ being used.
Thus a range of physical particle radii $R$ from \SI{25}{nm} to \SI{200}{nm} are resolved, with \SI{50}{nm} and \SI{100}{nm} being represented by two different combinations of cells and $\Delta x$.

\begin{figure}
\centering
      \includegraphics[width=\textwidth]{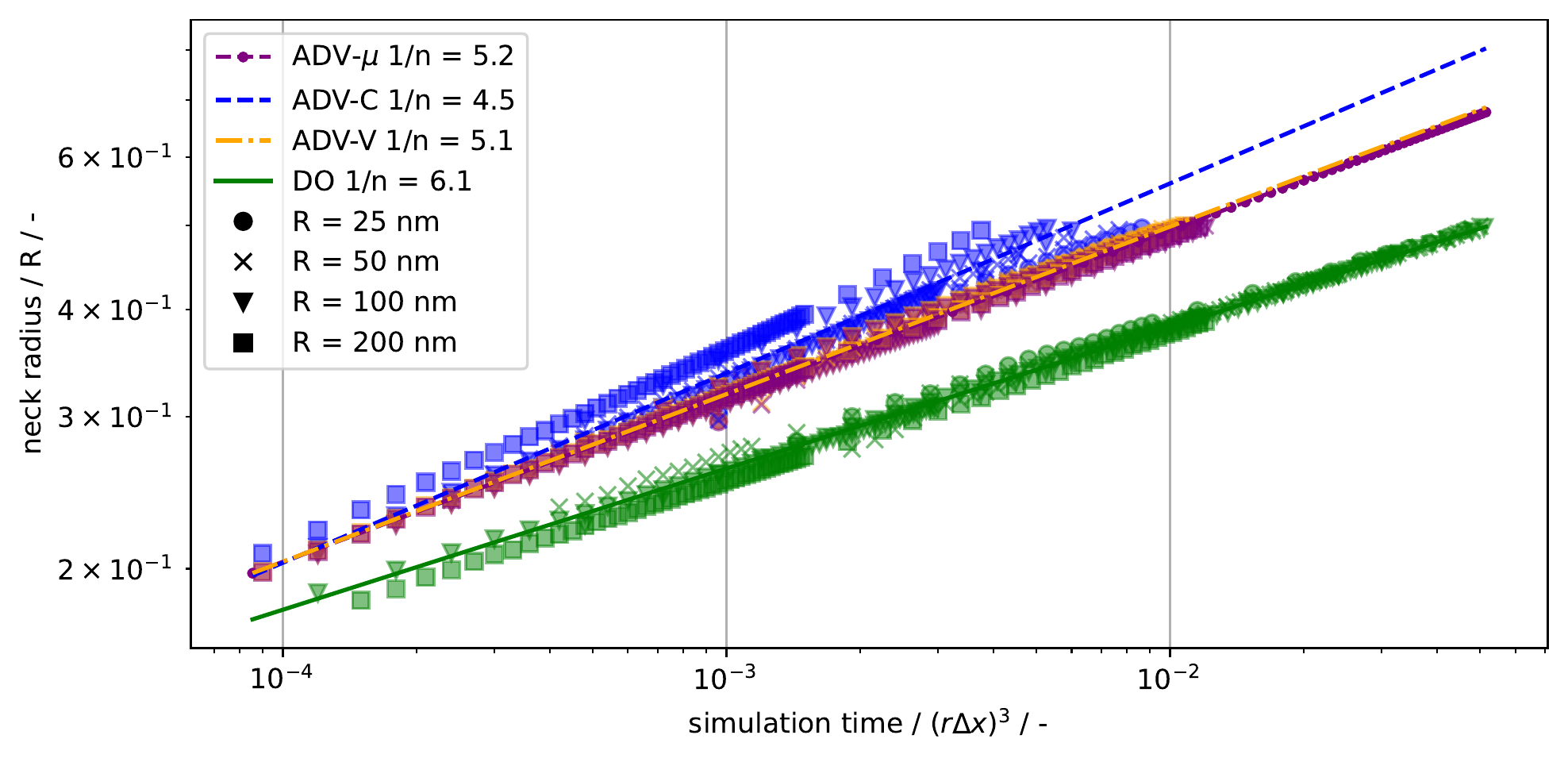}
      \caption{Neck size evolution up to $X/R=0.5$, with the time scaled following Herring's scaling law.
        The ADV models' exponent $1/n$ clusters in the expected range of 4-5, but the DO model shows an unexpected value of 6.
  All models except for ADV-C scatter closely and randomly around their master curve and thus follow Herring's scaling law.
  Model ADV-C tends to scatter upwards as particle size is increased.
      }
      \label{fig:neckgrowth}
\end{figure}

The time evolution of the relative neck radius $X/R$, scaled according to Herring's scaling law, is shown in \cref{fig:neckgrowth}.
The data is filtered such that a parabolic profile in the chemical potential is present within the grain boundary and for $X/R < 0.5$.
The former ensures that the simulation matches the theoretical expectation and that the interface is well-developed.
The latter excludes the approach to equilibrium, which the scaling laws do not represent and thus there is no sense in including that regime.
The regime is taken to be larger than the usual $X/R < 0.3$, as \cite{Parhami1999} still observed quite close matching up to $X/R=0.5$ for a similar dihedral angle.
As expected, the DO model shows the slowest evolution, whereas ADV-C shows the quickest evolution.
There is little difference in the evolution between the ADV-V and ADV-$\mu$ models, though as seen earlier, different equilibria will be obtained.
Models excluding ADV-C show mostly random scattering around their respective master curve, regardless of the chosen particle radius $R$.
For model ADV-C, the line tends to move upwards as the particle size is increased.
Thus a fixed choice of $c^{eq}_{gb}$ might not follow Herring's scaling law, though the present set of simulations allows no conclusive decision.
Furthermore, the slopes of curves differ from the classical two-particle model expectation of $1/5$\cite{Rahaman2003}.
The deviation is of similar magnitude as observed by other phase-field models of sintering\cite{Wang2006,Biswas2016}.
Interestingly, the present DO model seems to replicate the observed $n\sim1/6$ of \cite{Wang2006} rather closely, whereas the models including advection hit much closer to the expected $n=1/5$.
It might be that the factors employed by \cite{Wang2006} led to an evolution which was more dominated by diffusion rather than advection.

The effect of a change of $\Delta x$ while keeping the physical radius $R$ constant is that the curve is moved upwards, especially for shorter simulation times.
Excepting model ADV-C, these simulations approach each other for larger times and thus the size dependence should be completely included.

\begin{figure}
\centering
    \includegraphics[width=\textwidth]{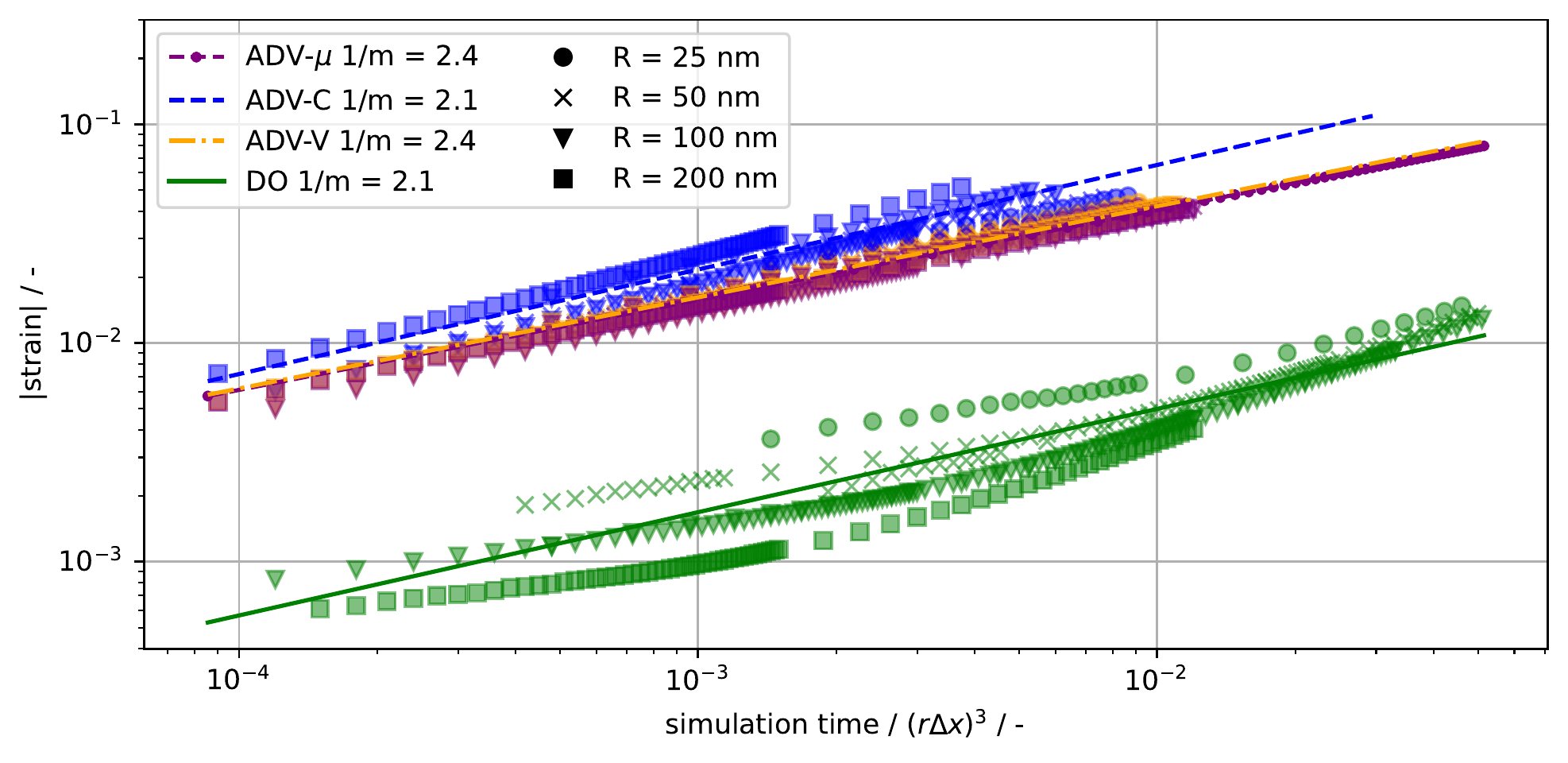}
  \caption{
  Absolute strain $|e|$ up to $X/R=0.5$.
  The expectation that $1/m=2/n$ for the strain is roughly observed, with the strain values for the DO model probably being too small to allow for a trustworthy evaluation.
  The upwards drift of model ADV-C with increasing particle size is observed again.
  }
  \label{fig:strain}
\end{figure}

The absolute strain $|e|$ is shown in \cref{fig:strain}.
The expectation that the observed exponent is half that of the neck growth law is roughly confirmed for the models with advection.
A similar deviation from Herring's scaling law is observed for model ADV-C.
Model DO tends to scatter strongly, likely due to its small amount of strain in the first place, so the value of the fitted exponent is likely wrong.
The equilibrium strain ($~0.333$ for models DO and ADV-$\mu$, $0.366$ for ADV-V and $0.429$ for ADV-C) could be observed for the simulations from the first study.
This is below the strain predicted by Kellet\cite{B.J.Kellet1989} for an infinite chain of cylinders, as also observed in \cite{Seiz2022}, and is likely explainable by finite size effects.

In total, the model ADV-$\mu$ seems to produce the most sensible results and thus will be employed in the next study.




\subsection{Large-scale three-dimensional simulations}
In \cite{Seiz2022} a small but persistent decrease in densification rate is observed as the number of particles in a chain is increased.
Since the chain geometry is quite restricted in its movement and does not contain porosity to fill, a small number of large-scale 3D simulations will be conducted to probe this effect further.
The initial conditions are generated by employing \cite{LAMMPS} as a discrete element simulation tool.
A periodic box of fixed size is filled with spheres of uniform size with a random velocity distribution, followed by letting the system evolve in an NVE ensemble while accounting for the translational and rotational degrees of freedom of the three-dimensional particles.
The particle interaction is described with a Hertzian contact law.
The resulting packing is then sliced to various extents, with larger slices always containing the smaller slices as subdomains.
The cuboid slices will be of size $c^3$ with $c \in \{200, 400, 800\}\si{nm}$, which with $\Delta x =0.1$ corresponds to domain sizes of $200^3$, $400^3$ and $800^3$ cells respectively.
The simulation volume $c^3$ will henceforth be used directly as a simulation label.
The individual particles are resolved with a radius of $12$ cells ($R=\SI{12}{nm}$), ensuring that there are bulk cells for each particle while allowing a large number of particles to be contained within the simulation domain.
A particle is only voxelized into the domain if its outer edge is at least 15 cells from the global boundary in order to exclude boundary effects from the phase-field.
This results in 263, 3446, and 34460 particles for the $200^3$, $400^3$, and $800^3$ domains respectively.
No-flux conditions are applied on all boundaries for all fields.
Each simulation is preprocessed by running the DO model for 5000 steps with equal bulk and vapor diffusivities of $D=\SI{1e-12}{m^2/s}$.
This is done to ensure that interfaces have already been established, as to reduce the influence of the grain boundary filtering function $g$ on the initial evolution.
After this step, all simulations are run with the parameters listed in \cref{tab:params} for at least an initial run of $300\,000$ time steps, with more depending on the observed evolution.
Grain growth is mostly suppressed by the choice of a small grain boundary mobility.
For the longest-running simulation, the mean grain size changed from $\SI{11.84}{nm}$ to $\SI{13.95}{nm}$, with less change for simulations running for a shorter time.
Given the small change in grain size, its effect on the density evolution should be negligible compared to other effects present.

\begin{figure}
\includegraphics[width=\textwidth]{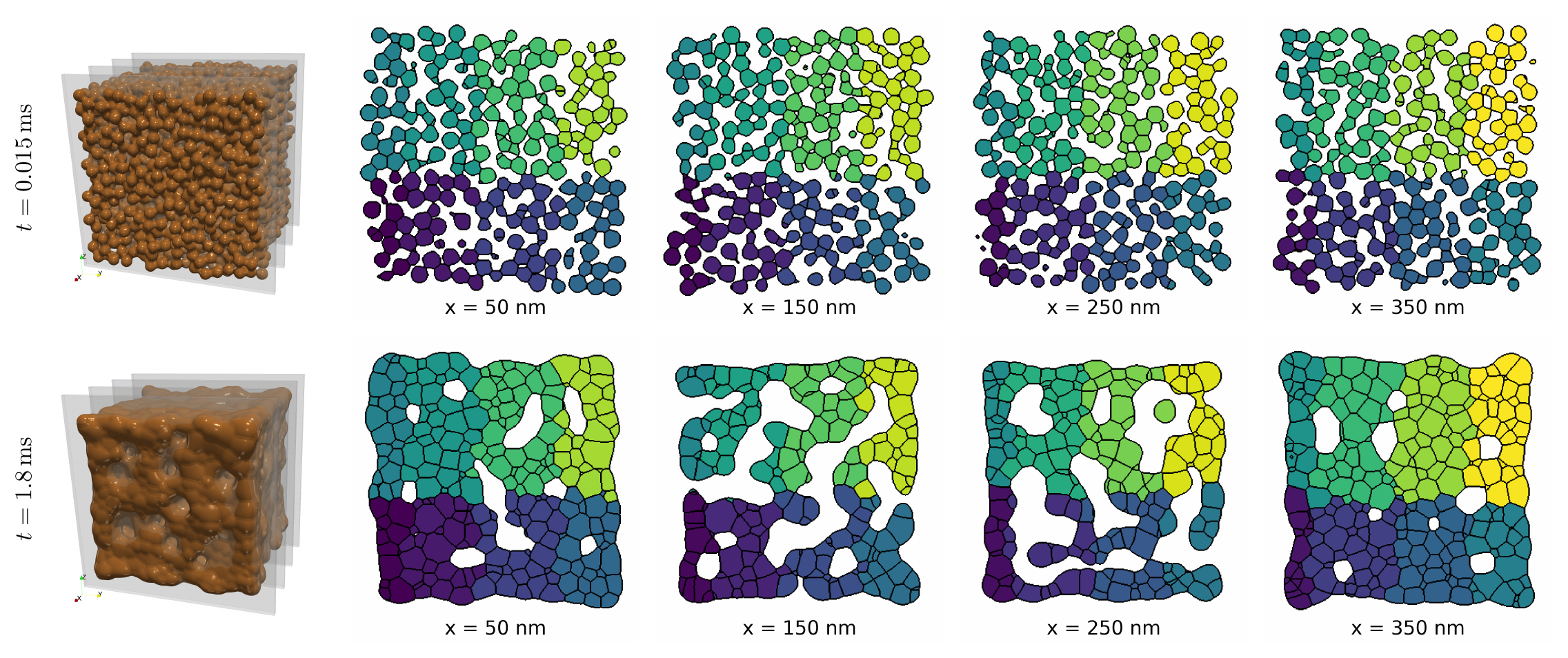}
\caption{The 3D green body as well as 2D slices of the $400^3$ domain calculated with model ADV-$\mu$ close to simulation start and at simulation end are shown.
The slice positions are indicated with the transparent planes.
Within the 2D slices, the surrounding vapor is depicted as white, any interfaces as black and the grain number with a colormap without physical meaning.
While initially the structure is homogeneous, as time progresses the outer edges become denser than the inner part of the green body.}
\label{fig:surf-pack400}
\end{figure}

Exemplarily, the surface of the structure at simulation start and simulation end for the $400^3$ domain is shown in \cref{fig:surf-pack400} along with 2D slices through the domain showing the grain structure.
While there is significant neck growth, barely any movement inwards is observed.
Furthermore, the 2D slices reveal that the inner part of the green body densifies much less quickly than the outer parts.
It should be noted that the entire green body stays connected during the process; videos of complete scans through the green body are deposited with the Supplementary Material.

\begin{figure}
\centering
 \includegraphics[width=\textwidth]{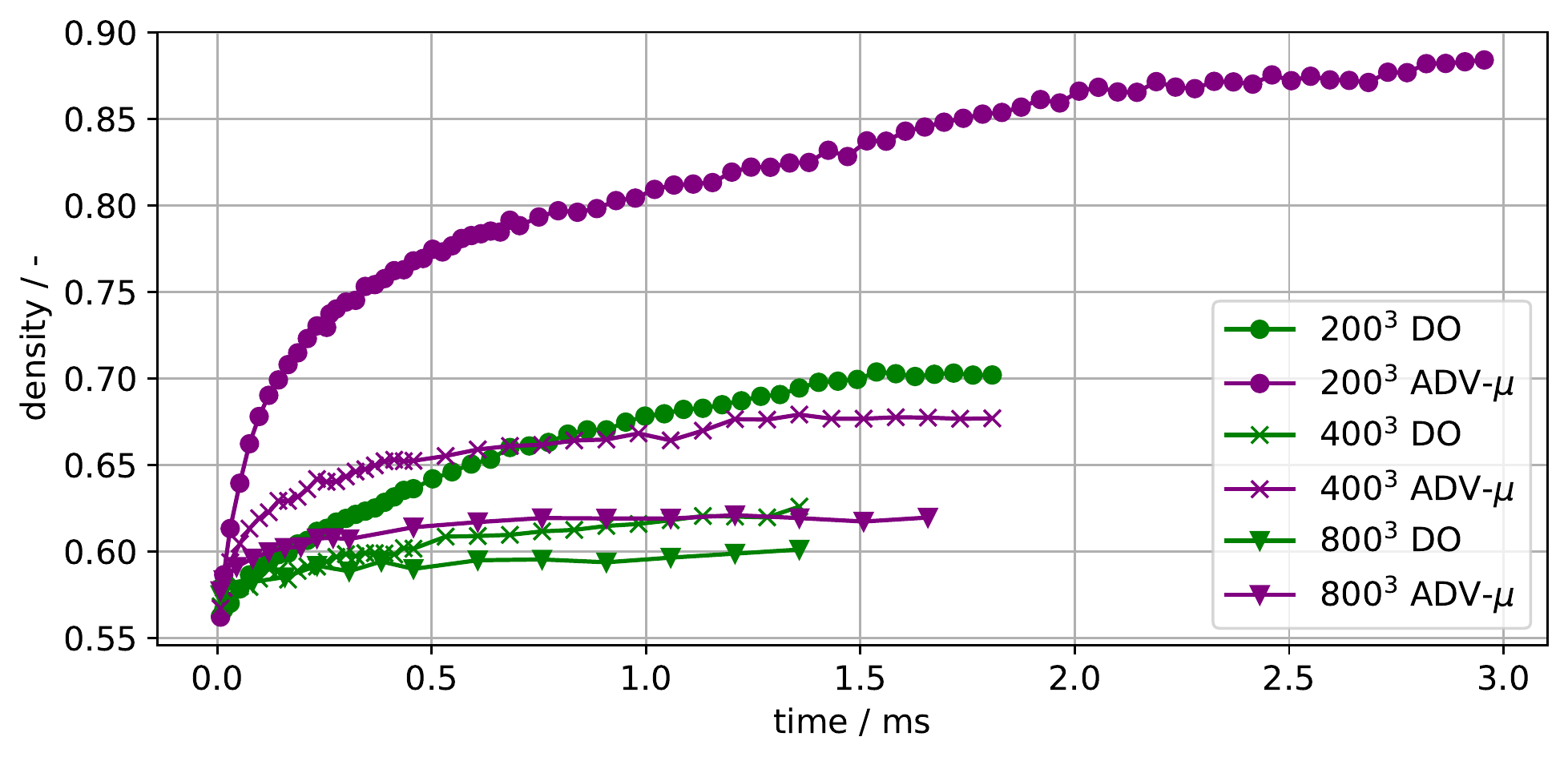}
 \caption{Density of the green bodies over time, for models DO and ADV-$\mu$ and various packing sizes.
 While model ADV-$\mu$ does densify more quickly than model DO, its densification rate is also strongly dependent on the system size.}
 \label{fig:3d-densify}
\end{figure}

The density evolution observed for this study is shown in \cref{fig:3d-densify}.
It can easily be seen that the DO model has a strong dependence of its densification on the green body size.
Furthermore, while the ADV-$\mu$ model does densify more quickly, it also has a strong dependence on the green body size.

\begin{figure}[h]
\begin{center}

 \begin{subfigure}[b]{0.45\textwidth}
 \includegraphics[width=\textwidth]{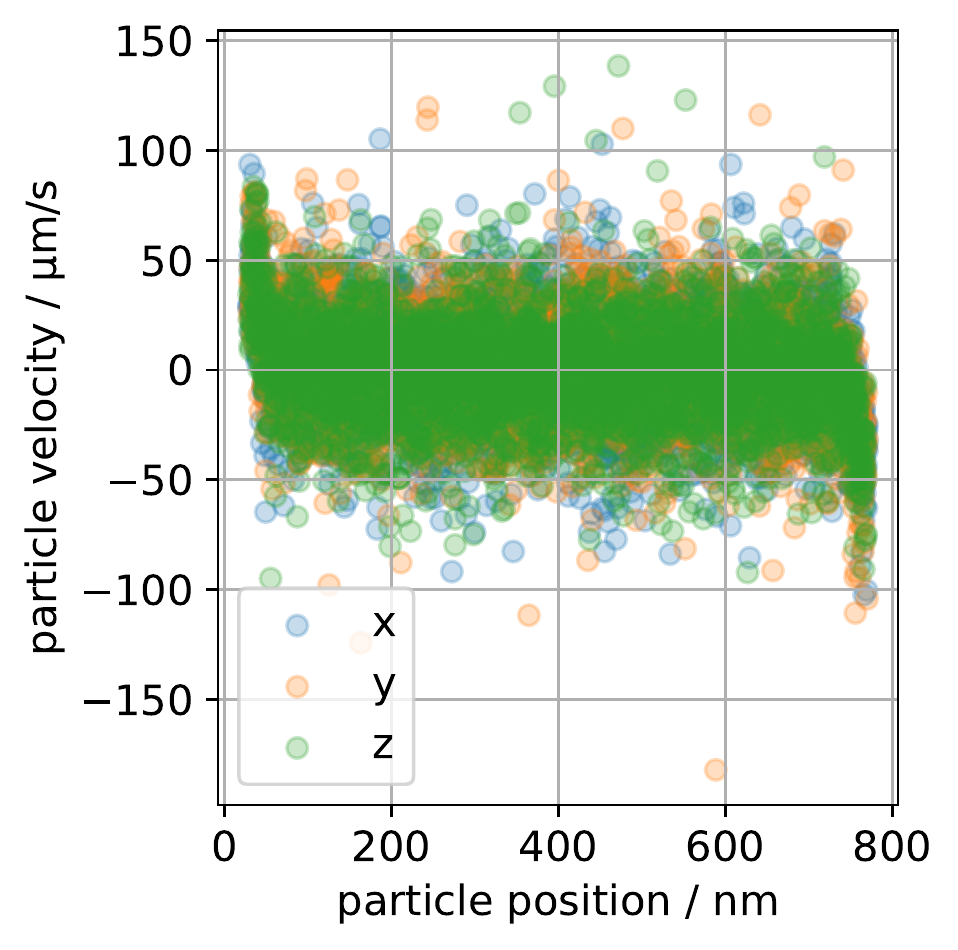}
    \caption{t = $\SI{0.045}{ms}$}
    \label{fig:3d-vels-early}
  \end{subfigure}
  \begin{subfigure}[b]{0.45\textwidth}
 \includegraphics[width=\textwidth]{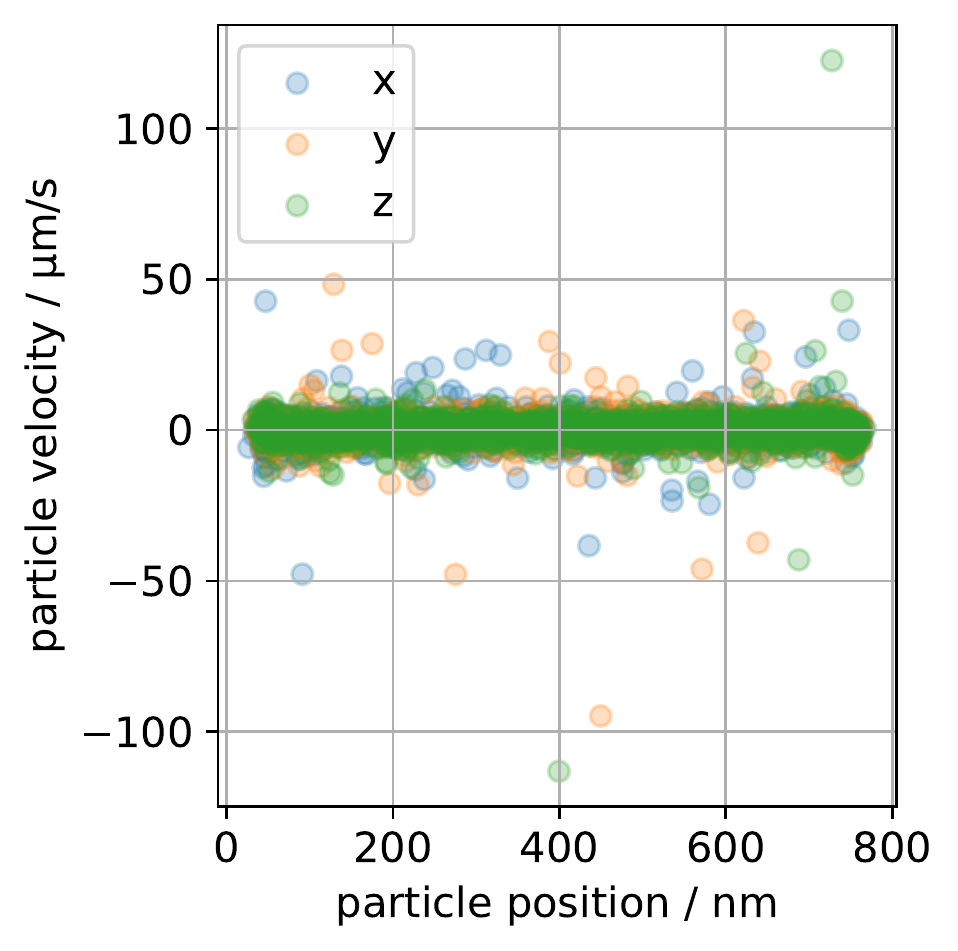}
    \caption{t = $\SI{1.4}{ms}$}
    \label{fig:3d-vels-later}
  \end{subfigure}
 \caption{
 Particle velocities over their respective barycenters for the $800^3$ domain and model ADV-$\mu$ for two times.
 The outer particles have significantly larger velocities, whereas any correlation between position and velocity is lost within the green body proper.
 This decorrelation becomes more pronounced as the simulation progresses, with local interactions causing high individual particle velocities.
 }
 \label{fig:3d-vels}
 \end{center}
\end{figure}

Thus the hypothesis stated in \cite{Seiz2022} is confirmed, in that the model for calculating advection velocities is lacking a part which eliminates this dependence.
The most relevant quantity to observe here is the spatial distribution of velocities.
Densification in principle means the reduction of occupied volume.
In the language of continuum mechanics, this is nothing more than demanding that the dilatation $\delta = \frac{\Delta V}{V_0} = tr(e)$ is negative, with the trace of the strain tensor $e$.
Differentiating this by time yields the same property for the strain rate tensor and its trace $tr(\dot{e}) = \nabla \cdot v$ which ought to be negative for densification to take place.
Thus for any control volume to densify, its $\nabla \cdot v$ needs to be negative.
Note that this should hold for macroscopic control volumes containing multiple particles.
It does not need to hold on a local basis, as e.g. $\nabla \cdot v$ is zero everywhere within the bulk of the particles due to the rigid body assumption.

Thus in order for a body to densify uniformly, its strain rate needs to be homogeneous, suggesting that its velocity is a linear function of position.
Of course, if a green body were nonhomogeneous in its vacancy absorption rate, this need no longer hold.
In the present case however all properties are isotropic and homogeneous to the extent that the structure is homogeneous; thus there is little reason for a deviation from linearity.
The velocity components are depicted over their particle's spatial coordinates in \cref{fig:3d-vels} within the $800^3$ domain, for every 10th particle.
Given the above discussion of the relationship between densification and velocity distribution, it is obvious that the present model will preferentially densify the outer edges, with the inner part showing almost no densification.
This is indeed observed as shown in \cref{fig:surf-pack400}.
Due to this non-uniform densification, no RVE can be found for this model, as the controlling parameter for the density evolution is now the ratio of inner particles to outer particles, which will not converge for finite domain sizes.

It should be noted that this conclusion is independent of how $c^{eq}_{gb}$ is determined, as the particle velocity will still only depend on local interactions.
A similar thought experiment as \cite{Seiz2022} conducted, for why a diffusion-only model fails to scale correctly with the number of particles in a chain, shows this easily:
Consider a control volume of sufficient size to be considered homogeneous on the inside of the packing:
Since only local interactions are taken into account and it is homogeneous, neighboring control volumes will have a similar magnitude and sign of the velocity.
Thus for neighboring control volumes there is little to no velocity gradient, which implies \emph{little to no densification} as per the above discussion.
Now consider a control volume on the edge of a packing:
Since the particles there have missing neighbours, they will have a significant nonzero velocity gradient to any control volume only containing inner particles and thus can densify w.r.t the inner control volumes.
What this implies is that the outer particles are \emph{implicitly} treated differently from the inner particles, which is the origin of the nonhomogeneous densification.
The only feature of the model necessary to arrive at this conclusion is that only local interactions enter the velocity calculation for a fixed time, and thus the precise value of $c^{eq}_{gb}$ is irrelevant.

\section{Conclusion}
In the present paper, a previously published phase-field model of sintering is extended with advective terms in order to better represent densification during sintering.
Several key insights of a recent work on the calculation of rigid body velocities for sintering were incorporated as to ensure consistency with the free energy functional, resulting in multiple potential models.
The new models are compared by testing the free energy evolution, the equilibrium state, as well as their dynamic evolution.
It is observed that among the advective models, only ADV-$\mu$, which estimates the grain boundary equilibrium density by averaging the surface chemical potential, is consistent with the free energy functional.
All advective models roughly reproduces the expected scaling laws of the neck size with time, both in terms of the time and particle size dependence.
Furthermore, the approach of centers as quantified by the strain is observed to also reproduce the expected scalings.
Based on these results, the most promising model ADV-$\mu$ is employed in order to simulate large scale 3D structures in order to seek representative volume elements.
However, it is observed that even this model shows a strong dependence of the densification on the green body size and thus no RVEs could be identified.
The spatial distribution of the velocity is identified as the likely origin of this dependence.
Future work will focus on eliminating this dependence as to allow identification of RVEs and produce a quantitative phase-field model of sintering.

%

\section*{Supplementary Material}
The Supplementary Material of this paper is available at {\url{https://doi.org/10.5281/zenodo.7755462}}.

\section*{Acknowledgements}
This work was partially performed on the national supercomputer Hawk at the High Performance Computing Center Stuttgart (HLRS) under the grant number \hl{pace3d}.
The authors gratefully acknowledge financial support for the modelling of sintering by the \hl{DFG under the grant number NE 822/31-1 (Gottfried-Wilhelm Leibniz prize)} and support for the parallelization and code optimization by \hl{KNMFi within the programme MSE (P3T1) no. 43.31.01}.


\bibliography{literatur}

\end{document}